# Investigation of Dosimetric Parameters of $^{192}$Ir MicroSelectron v2 HDR Brachytherapy Source Using EGSnrc Monte Carlo Code

Hamza Naeem[1, 2, 3], Chaobin Chen[2,3]*, Huaqing Zheng[2,3], Ruifen Cao[2, 3], Xi Pei[2, 3], Liqin Hu[1, 2, 3], Yican Wu[1, 2, 3]

(1. School of Nuclear Science and Technology, University of Science and Technology of China, Hefei, Anhui, 230027, China
2. Key Laboratory of Neutronics and Radiation Safety, Institute of Nuclear Energy Safety Technology, Chinese Academy of Sciences, Hefei, Anhui, 230031, China
3. Collaborative Innovation Center of Radiation Medicine of Jiangsu Higher Education Institutions, Suzhou 215006, China)
*Corresponding Author, Email: chaobin.chen@fds.org.cn

**Abstract**
The $^{192}$Ir sources are widely used for high dose rate (HDR) brachytherapy treatments. The aim of this study is to simulate $^{192}$Ir MicroSelectron v2 HDR brachytherapy source and calculate the air kerma strength, dose rate constant, radial dose function and anisotropy function established in the updated AAPM Task Group 43 protocol. The EGSnrc Monte Carlo (MC) code package is used to calculate these dosimetric parameters, including dose contribution from secondary electron source and also contribution of bremsstrahlung photons to air kerma strength. The Air kerma strength, dose rate constant and radial dose function while anisotropy functions for the distance greater than 0.5 cm away from the source center are in good agreement with previous published studies. Obtained value from MC simulation for air kerma strength is $9.762 \times 10^{-8}$ UBq$^{-1}$ and dose rate constant is $1.108\pm0.13\%$ cGyh$^{-1}$U$^{-1}$.

## 1. Introduction

Brachytherapy is the placement of sealed radionuclides close to the surface to be treated so that SSD is comparable to treatment depth, resulting in a rapid dose fall off. This is achieved by (i) placing the source directly onto the surface to be treated (molds, plaques, beta applicators) (ii) inserting the sources into body cavities (intra-cavity) (iii) implanting the source, temporarily or permanently, directly into the tumor (interstitial).

$^{192}$Ir high dose rate (HDR) brachytherapy sources are commonly used in the treatment of tumors of the cervix, breast, prostate, lungs and others[1]. Clinically theses sources require an accurate determination of all those dosimetric parameters which are used in treatment planning system (TPS) [2]. American Association of Physicists in Medicine (AAPM) recommended in task group reports 43 and 43U1 (TG-43, TG43U1) [3,4], accurate dosimetric data must be acquired as input for TPS on realistic geometry and on mechanical characteristics of the sources by standard methods, either from Monte Carlo simulations or from experimental measurements. MC calculations plays an important role to provide dosimetrical data by modeling the geometry of the radioactive sources and interaction process of particles emitted in a decay process. Commercially MC codes are widely available. These codes track the history of particle and allow the user to obtain data from the points where experimental measurements is very difficult. Dose distribution close to radioactive source (<0.2cm) is rarely known in intravascular radiotherapy. In traditional brachytherapy, the effects of low energy photons and secondary electrons are ignored at near source region. Wang and Li [5] investigated that path length estimator is not accurate at distance below 0.2cm near the source due to electronic disequilibrium and contribution of beta spectrum of $^{192}$Ir on dose. A number of studies regarding near source region for high energy brachytherapy source suggested that electronic disequilibrium exists at distance below 0.2 cm and it should not be ignored [6-8].

In this paper, MC calculation results of photon spectrum in air and dose distribution in water of $^{192}$Ir microSelectron v2 HDR brachytherapy source are presented. We investigate the effect of beta spectrum (average energy of ~181 keV) on dose and contribution of bremsstrahlung photons from secondary electron induced by decay gamma photons from $^{192}$Ir, using FLURZnrc and DOSRZnrc user codes in the EGSnrc system. The EGSnrc codes are modified version of EGS4, which is a general purpose package for Monte Carlo simulation, capable for accurate simulation of electron and photon through a random process based on physical and statistical principles of particle transport and interaction [9-11]. Fluence data results are compared with Borg and Roggers [12], and dosimetric parameters are compared with Daskalov [13] and with the data base provide by Taylor and Rogers [14]. This research was based on Advance/Accurate Radiotherapy System (ARTS), developed by FDS team [15-18].

## 2. Material and Methods

### 2.1. Radioactive source description

The details of geometry design and material composition of the microSelectron v2 HDR $^{192}$Ir source used in our Monte Carlo study are taken from Borg and Rogers [12]. The active core of pure $^{192}$Ir metal cylinder is 0.36 cm in length and 0.065 cm in diameter. Uniform exposure is expected to be distributed around it. Around the core is encapsulation of AISI 316L (stainless steel) with 0.45 cm length and 0.09 cm outer diameter, and linked to a 0.02 cm long steel cable having 0.07 cm diameter.

**Table 1**: Composition and density of the materials used in Monte Carlo calculation.

| Material | Element | Composition [% by weight] | Density [g cm$^{-3}$] |
|---|---|---|---|
| Air (dry, near sea level) | C<br>N<br>O<br>Ar | 0.0124<br>75.5267<br>23.1781<br>1.2827 | 1.205E-03 |
| Iridium | Ir |  | 22.42 |
| Steel (AISI 316) | Si<br>Cr<br>Mn<br>Fe<br>Ni<br>Mo | 7<br>17<br>1<br>66.8<br>12<br>25 | 8.06 |
| Steel cable (AISI 302) | C<br>Si<br>Cr<br>Mn<br>Fe<br>Ni | 1<br>7<br>18<br>10<br>71.2<br>9 | 8.06 |

Figure 1a illustrates the actual model of microSelectron HDR $^{192}$Ir source and Figure 1b shows the source model used in our Monte Carlo calculations. Our model is only an approximation to actual geometry, neglecting the round shape because the user codes FLURZnrc and DOSRZnrc are based on RZ (radius plane) coordinates, which is a limitation in the designing of the actual geometry of the tip of the source.

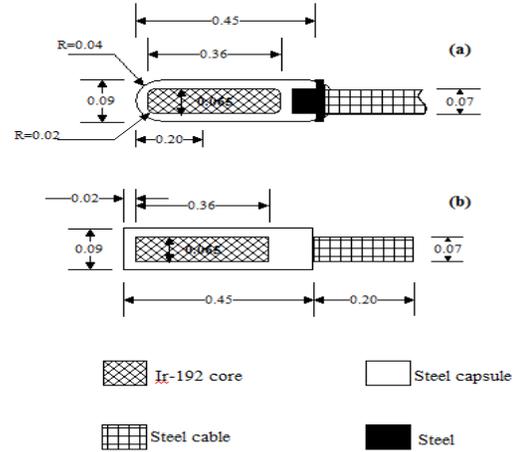

**Figure 1**: The microSelectron-HDR $^{192}$Ir source design. a) actual source model, b) source model used in this work [12].

The half-life of $^{192}$Ir is 73.825 days[19] and one average decay will emit one electron and 2.363 photons. The source activity as function of photons emitted per sec can be expressed as:

$$N_{photon} = A.(2.363 \pm 0.3\%) \quad [photons\ s^{-1}] \quad (1)$$

Where the uncertainty is calculated from the published spectrum of $^{192}$Ir by Duchemin and Coursol [20].

### 2.2. Monte Carlo Simulation

To calculate air kerma strength per activity and dose rate distribution in water in terms of $\mu Gym^2h^{-1}$, we use FLURZnrc and DOSRZnrc user codes respectively.

For photon transport, fluorescent emission of K-shell and L-Shell characteristic x-ray, Rayleigh and bound Compton scattering have been simulated in all regions. For electron transport, Bremsstrahlung production, multiple scattering as well as Moller and Bhabha scattering have been taken into account. The Brems cross-sections are taken from NIST database [21] while photon cross-section are taken from PEGS4 cross-section library[22]. The bare $^{192}$Ir photon spectrum needed for Monte Carlo calculation was taken from Duchemin and Cursol [19].

### 2.3. Calculation of Photon fluence spectrum and air kerma strength

FLURZnrc user code was used for calculation of photon fluence spectrum and air kerma strength. The fluence was calculated in vacuum at 5 cm distance from the source along the transverse axis at 5 keV bins spectrum in fluence per MeV units was modify to that per decay and compared with the spectrum calculated by Borg and Rogers [12].

The air kerma strength, $S_k$, measures the strength of brachytherapy source, which was first introduced by AAPM task group report number 32[23]. For convenience, it is denoted by U where $1U = 1\mu Gym^2h^{-1} = 1cGycm^2h^{-1}$ and numerically identical to reference air kerma strength recommended by ICRU 38 [24] and ICRU 60[25]. Air kerma strength is defined as the product of air kerma rate in free space at a distance d, $\dot{K}_{air}(d)$ and square of the distance, $d$:

$$S_k = \dot{K}_{air}(d) \cdot d^2 \qquad (2)$$

The air kerma strength per unit source activity, $S_k/A$, was calculated at the distance of 1, 2, 5, 10, 20 and 50 cm distance from the center of the source at transverse axis, in terms of $UBq^{-1}$.

The air kerma per initial particle was calculated from fluence differential in energy per initial particle in 5 keV bins and mass energy-absorption coefficients for dry air in the middle of each bin, by using the following equation:

$$S_k/A = 3.6 \times 10^9 \times 1.602 \times 10^{-10} \times 2.363 \times \sum_{5keV}^{Emax} \phi(E_i) \cdot \left(\frac{\mu_{en\,(E_i)}}{\rho}\right) \cdot \Delta E \qquad [UBq^{-1}] \qquad (3)$$

Where $E_i$ is the mid-point of each energy bin and $\Delta E$ is the bin size. $\left(\frac{\mu_{en\,(E_i)}}{\rho}\right)$ is x-ray mass energy-absorption coefficient at energy $E_i$ which was taken from NIST compilation [23]. $\phi(E_i)$ [$MeV^{-1}cm^{-2}$] is the photon fluence differential in energy within each energy bin. The factor $1.602 \times 10^{-10}$ is required to convert $K_{air}$ from MeV $g^{-1}$ to Gy. 2.363 is taken from the above equation (1). The factor $3.6 \times 10^9$ is used to convert the unit $Gym^2s^{-1}Bq^{-1}$ to $UBq^{-1}$.

In the direction of transverse axis, the scoring regions for fluence are 0.01 cm thick air cylinder and 1 cm distance and 0.05 cm thick air cylinder. The length of the scoring region is 0.02 cm which is centered at mid of the active length. The other parameters used are: Number of histories= $10^9$, ECUT=2.0 MeV, PCUT=0.001 MeV, ESTEPE=0.25.

### 2.4. Dose calculation in water phantom

Dose calculations were carried out with the source placed at the center of a water phantom using DOSRZnrc user code. Water phantom is a cylinder of 15 cm radius and 30 cm of height. Dose distributions surrounding the source were calculated within a matrix of two dimensional grid points perpendicularly away and along both directions at a distance of 0.00, 0.10, 0.25, 0.35, 0.5, 0.75, 1.00, 1.50, 2.00, 3.00, 5.00, 7.00 cm from the source center. Scoring regions were 0.05 cm in r and 0.02 cm in z-axis for distance less than 3 cm and 0.1cm in r and 0.05 cm in z-axis direction for distance greater than 3 cm. Input parameters used are: number of histories=$10^{11}$ (to maintain uncertainty under 0.5%), ECUT=0.521 MeV, PCUT=0.001 MeV, ESTEPE=0.25. No variance reduction technique was implemented.

### 2.5. TG-43 dosimetric parameters

The dosimetric parameters calculated around microSelectron v2 HDR brachytherapy source are dose rate constant ($\Lambda$), geometry function $G_L(r,\theta)$, radial dose function $g_L(r)$ and anisotropy function $F(r,\theta)$ as described by TG43[3].

### 3. Results

### 3.1. Photon fluence spectrum

The photon fluence spectrum in 5 keV bins of $^{192}$Ir micoSelectron HDR v2 source in air at 5 cm in the direction of transverse axis of the source is shown in figure 3 and compared the spectrum calculated by Borg and Roger's [12].

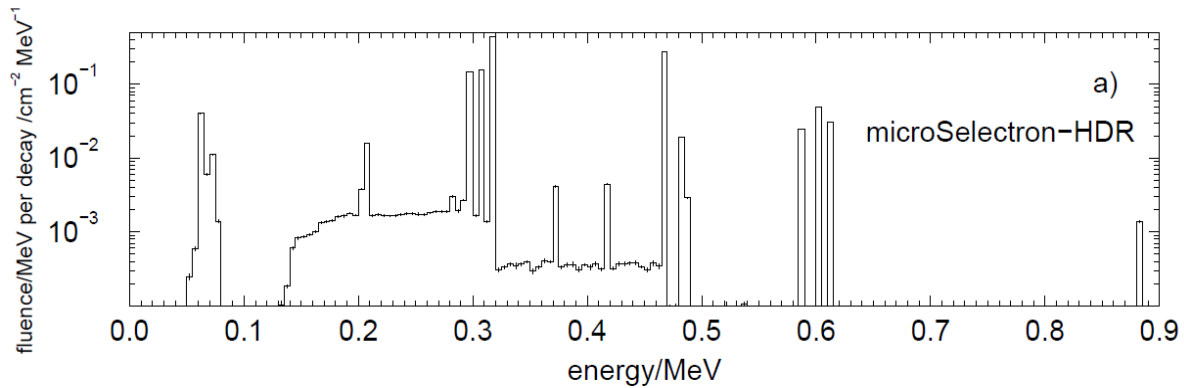

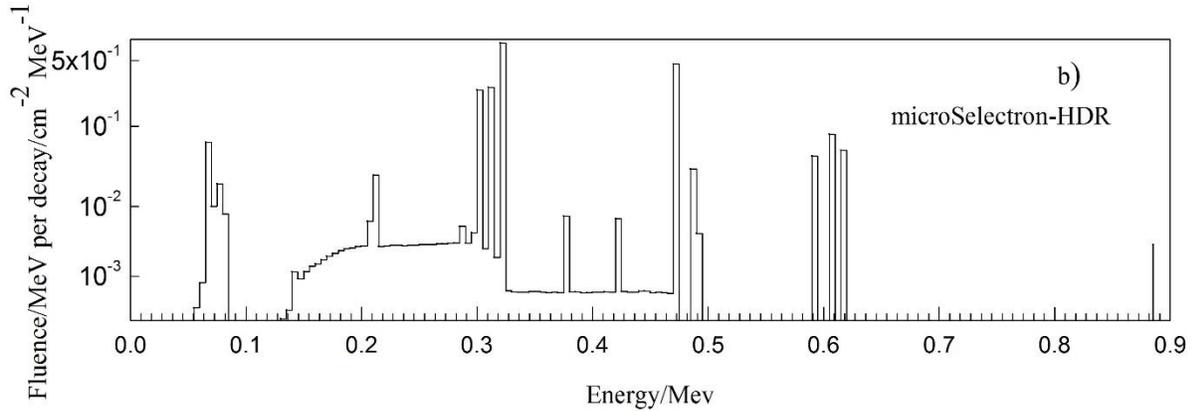

Figure 2: a) spectrum calculated by Borg and Rogers [12] b) spectrum from our study.

### 3.2. Air Kerma Strenght

The value of air-kerma strength per unit source activity by using FLURZnrc was $9.762 \times 10^{-8}$ UBq$^{-1}$ which is very close to $9.73 \times 10^{-8}$ UBq$^{-1}$ obtained by Borg and Roger's [12]. Bremsstrahlung contribution was also included in this work which increase $S_k/A$ values 0.2% for our $^{192}$Ir microSelectron HDR source.

### 3.3. Dose rate constant

The dose rate constant ʌ for $^{192}$Ir HDR source in water phantom was 1.109±0.3% cGyh$^{-1}$U$^{-1}$ .by using DOSRZnrc. This value is very close to 1.108±0.13% cGyh$^{-1}$U$^{-1}$ obtained by Daskalov[13]and 1.109±0.20% obtained by Taylor and Roggers[14].

Detailed dose rate per unit air kerma strength (cGyh$^{-1}$U$^{-1}$) is shown in table 2.

**Table 2:** Dose rate per unit air kerma strength (cGyh$^{-1}$U$^{-1}$) of $^{192}$Ir microSelectron v2 HDR source

| z (cm) | y (cm) | | | | | | | | | | |
|---|---|---|---|---|---|---|---|---|---|---|---|
| | 0.00 | 0.10 | 0.25 | 0.50 | 0.75 | 1.00 | 1.50 | 2.00 | 3.00 | 5.00 | 7.00 |
| **7.00** | 0.016 | 0.016 | 0.016 | 0.016 | 0.017 | 0.016 | 0.016 | 0.016 | 0.016 | 0.012 | 0.009 |
| **6.00** | 0.021 | 0.022 | 0.022 | 0.023 | 0.023 | 0.023 | 0.023 | 0.023 | 0.021 | 0.016 | 0.011 |
| **5.00** | 0.031 | 0.032 | 0.032 | 0.033 | 0.032 | 0.034 | 0.034 | 0.033 | 0.029 | 0.020 | 0.013 |
| **4.00** | 0.048 | 0.049 | 0.049 | 0.052 | 0.052 | 0.052 | 0.053 | 0.049 | 0.041 | 0.025 | 0.015 |
| **3.00** | 0.084 | 0.084 | 0.087 | 0.092 | 0.094 | 0.095 | 0.088 | 0.079 | 0.059 | 0.031 | 0.018 |
| **2.50** | 0.113 | 0.120 | 0.127 | 0.133 | 0.136 | 0.133 | 0.119 | 0.103 | 0.070 | 0.034 | 0.019 |
| **2.00** | 0.183 | 0.186 | 0.197 | 0.211 | 0.212 | 0.199 | 0.166 | 0.134 | 0.083 | 0.037 | 0.020 |
| **1.50** | 0.322 | 0.331 | 0.366 | 0.379 | 0.356 | 0.317 | 0.238 | 0.176 | 0.098 | 0.040 | 0.020 |
| **1.00** | 0.740 | 0.775 | 0.858 | 0.811 | 0.675 | 0.539 | 0.337 | 0.222 | 0.112 | 0.042 | 0.021 |
| **0.75** | 1.356 | 1.445 | 1.542 | 1.287 | 0.689 | 0.702 | 0.392 | 0.246 | 0.116 | 0.043 | 0.021 |
| **0.50** | 3.405 | 3.695 | 3.363 | 2.158 | 1.348 | 0.884 | 0.444 | 0.263 | 0.120 | 0.044 | 0.021 |
| **0.25** | 23.058 | 16.801 | 8.598 | 3.483 | 1.742 | 1.037 | 0.483 | 0.276 | 0.124 | 0.044 | 0.022 |
| **0.10** | - | 60.398 | 14.082 | 4.145 | 1.921 | 1.100 | 0.499 | 0.279 | 0.125 | 0.044 | 0.022 |
| **0.00** | - | 68.206 | 15.529 | 4.294 | 1.951 | 1.109 | 0.500 | 0.280 | 0.125 | 0.044 | 0.022 |
| **-0.10** | - | 63.221 | 14.527 | 4.126 | 1.920 | 1.094 | 0.499 | 0.283 | 0.124 | 0.044 | 0.022 |
| **-0.25** | - | 23.650 | 9.856 | 3.508 | 1.759 | 1.052 | 0.484 | 0.278 | 0.124 | 0.044 | 0.022 |
| **-0.50** | 3.160 | 3.689 | 3.433 | 2.138 | 1.352 | 0.887 | 0.445 | 0.265 | 0.121 | 0.044 | 0.022 |
| **-0.75** | 1.250 | 1.361 | 1.525 | 1.268 | 0.966 | 0.717 | 0.398 | 0.247 | 0.119 | 0.043 | 0.022 |
| **-1.00** | 0.661 | 0.725 | 0.840 | 0.802 | 0.676 | 0.545 | 0.340 | 0.224 | 0.112 | 0.042 | 0.021 |
| **-1.50** | 0.301 | 0.309 | 0.349 | 0.373 | 0.358 | 0.321 | 0.240 | 0.177 | 0.100 | 0.040 | 0.021 |
| **-2.00** | 0.170 | 0.175 | 0.191 | 0.208 | 0.210 | 0.202 | 0.169 | 0.135 | 0.085 | 0.037 | 0.020 |

| | | | | | | | | | | |
|---|---|---|---|---|---|---|---|---|---|---|
| **-2.50** | 0.113 | 0.111 | 0.119 | 0.129 | 0.136 | 0.135 | 0.121 | 0.104 | 0.072 | 0.035 | 0.019 |
| **-3.00** | 0.078 | 0.077 | 0.084 | 0.090 | 0.094 | 0.094 | 0.090 | 0.080 | 0.060 | 0.032 | 0.018 |
| **-4.00** | 0.045 | 0.046 | 0.047 | 0.050 | 0.052 | 0.053 | 0.053 | 0.050 | 0.041 | 0.025 | 0.015 |
| **-5.00** | 0.030 | 0.029 | 0.031 | 0.031 | 0.033 | 0.033 | 0.034 | 0.033 | 0.030 | 0.020 | 0.013 |
| **-6.00** | 0.020 | 0.021 | 0.020 | 0.022 | 0.022 | 0.022 | 0.023 | 0.023 | 0.022 | 0.016 | 0.011 |
| **-7.00** | 0.015 | 0.015 | 0.015 | 0.016 | 0.016 | 0.016 | 0.016 | 0.016 | 0.016 | 0.013 | 0.009 |

### 3.4. Radial dose function

The radial dose function $g_L(r)$ was calculated using line source approximation at 11 different radial distance ranging from 0.1 to 7 cm. Results are shown in figure 4 and compared with Daskalov [13] and Taylor and Rogers [14].

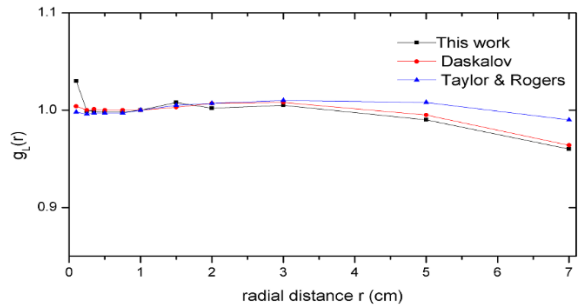

**Figure 3:** Radial dose function calculated in this work and compared with reference data.

### 3.5. Anisotropy function

Anisotropy function $F(r,\theta)$ (using line source approximation) values at 0.5, 1.0, 2.0, 3.0 and 5.0 cm from the source are shown in Figure 5 and compared with Daskalov [13] and Taylor and Rogers[14]. Good agreement obtained at distance greater than 0.5 cm.

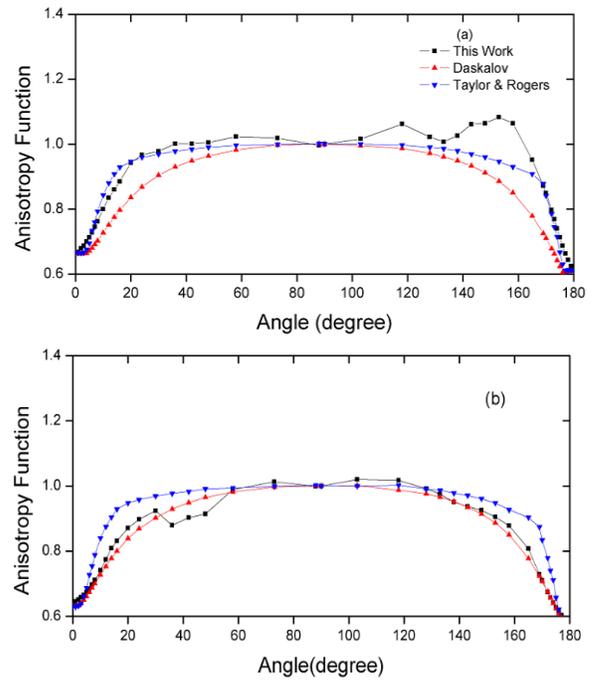

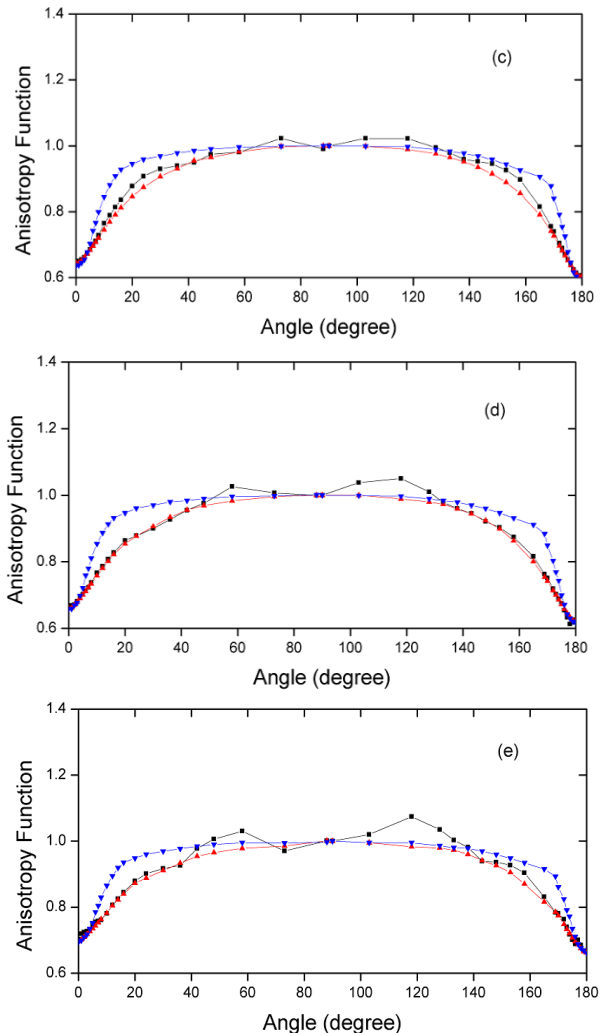

**Figure 4:** Anisotropy Function at radii (a) 0.5 (b) 1.0 cm (c) 2.0 cm, (d) 3.0 cm, (e) 5.0 cm.

### 4. Conclusion

In this work，EGSnrc user code was used to investigate dosimetric parameters of $^{192}$Ir microSelectron v2 HDR brachytherapy source and secondary electron effect on absorbed dose from $^{192}$Ir was also studied，it provides us accurate dosimetric parameters specially near to the source. Dose rate constant and radial dose function are in good agreement with the published data. For radii greater than 0.5 cm, anisotropy function also shows good results. At a short distance from the source i.e. within 0.25 cm and for specific polar angles ($<20^0$ and $>170^0$) from longitudinal axis, difference in TG-43 parameters occurs due to simplified source geometry (i.e. tips and encapsulation of the source). For accurate dosimetric parameters, we recommend to consider the effect of beta spectrum of $^{192}$Ir. This study demonstrated that Monte Carlo code DOSRZnrc is well suited for that.

### Acknowledgements

This work was supported by the Strategic Priority Research Program of Chinese Academy of Sciences (No. XDA03040000), the National Natural Science Foundation of China (No.11305205, No.11305203 and No.11405204), the National Natural Science Foundation of Anhui Provincial (No.1308085QH138 and No.1508085QH180), Anhui Provincial Special project for High Technology Industry and Special Project of Youth Innovation Promotion Association of Chinese Academy of Sciences. In addition, the authors would like to thank the great help from FDS Team.

### Reference

[1] E. Van Gerbaulet, Alain Pötter, Richard Mazeron, et al. The GEC ESTRO Handbook of Brachytherapy, 2002.

[2] R. Nath, L. L. Anderson, J. A. Meli, et al. Code of practice for brachytherapy physics: report of the AAPM Radiation Therapy Committee Task Group No. 56. American Association of Physicists in Medicine, Med. Phys, vol. 24, no. 10, pp. 1557–1598, 1997.

[3] M. J. Rivard, B. M. Coursey, L. a DeWerd, et al. Update of AAPM Task Group No. 43 Report: A revised AAPM protocol for brachytherapy dose calculations, Med. Phys, vol. 31, no. 3, pp. 633–674, 2004.

[4] R. Nath, L. L. Anderson, G. Luxton, et al. Dosimetry of interstitial brachytherapy sources: recommendations of the AAPM Radiation Therapy Committee Task Group No. 43. American Association of Physicists in Medicine, Med. Phys, vol. 22, no. 2, pp. 209–234, 1995.

[5] R. Wang and X. A. Li. Dose characterization in the near- source region for two high dose rate brachytherapy sources, Med. Phys, vol. 29, pp. 1678-1686, 2002.

[6] D. Baltas, P. Karaiskos, P. Papagiannis, et al. Beta versus gamma dosimetry close to Ir-192 brachytherapy sources, Med. Phys, vol. 28, pp. 1875-1882, 2001.

[7] P. Papagiannis, A. Angelopoulos, E. Pantelis, et al. Dosimetry comparison of $^{192}$Ir sources, Med.Phys, vol. 29, pp. 2239-2246, 2002.

[8] N. Patel, S. Chiu-Tsao, Y. Ho, et al. High beta and electron dose from $^{192}$Ir: implications for

'Gamma' intravascular brachytherapy, Int. J. Radiat. Oncol, Biol, Phys. vol. 54, pp. 1521-1528, 2004

[9] I. Kawrakow. The EGSnrc Code System : Monte Carlo Simulation of Electron and Photon Transport, NRCC Report PIRS-701, 2011.

[10] D. W. O. Rogers, I. Kawrakow, J. P. Seuntjens, et al. NRC User Codes for EGSnrc, NRCC Report PIRS-702(revC),2011.

[11] E. Mainegra-hing. User Manual for egs_inprz , a GUI for the NRC RZ user-codes, NRCC Report PIRS-801 (RevB), 2011.

[12] J. Borg and D. W. O. Rogers. Monte Carlo Calculations of Photon Spectra in Air from 192 Ir Sources, NRC Report PIRS-629r, pp. 0–37, 1999.

[13] G. M. Daskalov, E. Löffler, J. F. Williamson, et al. Monte Carlo-aided dosimetry of a new high dose-rate brachytherapy source. Med.Phys. 25,pp.2200-2208,1998.

[14] R.E.P.Taylor and D.W.O.Roggers. An EGSnrc Monte Carlo-calculated database of TG-43 parameters, Med.Phys. 35, pp. 4228-4241, 2008.

[15] Y. Wu, S. Gang, C. Ruifen, et al. Development of Accurate/Advanced Radiotherapy Treatment Planning and Quality Assurance System (ARTS), Chinese Physics C (HEP & NP), vol. 32(Suppl. II), pp. 177-182, 2008.

[16] C. Ruifen, Y. Wu, X. Pei , et al. Multi-objective optimization of inverse planning for accurate radiotherapy, Chinese Physics C, vol. 35, pp. 313-317, 2011.

[17] H. Zheng, G. Sun, G. Li, et al. Photon Dose Calculation Method Based on Monte Carlo Finite-Size Pencil Beam Model in Accurate Radiotherapy, Communications in Computational Physics, vol. 14(5), pp. 1415-1422, 2013.

[18] Y. Wu, J. Song, H. Zheng, et al. CAD-Based Monte Carlo Program for Integrated Simulation of Nuclear System SuperMC. Annals of Nuclear Energy, vol. 82, pp. 161-168, 2015

[19] B.Duchemin and N.Coursol. Reevaluation de I' [192]Ir. Technical Note LPRI/93/018, DAMRI,CEA,France. 1993.

[20] B.Duchemin and N.Coursol. LARA-90. LARA-LMRI-1990,DAMRI/LMRI, CEA,France. 1990.

[21] J. H. Hubbell and S. M. Seltzer. Tables of X-Ray mass attenuation coefficients and mass energy-absorption coefficients 1 keV to 20 MeV for Elements Z=1 to 92 and 48 Additional substances of Dosimetric Interest, Technical Report NISTIR 5632,NIST, Gaithersburg, MD 20889 (1995).

[22] J.H.Hubbell, Photon Cross Sections, Attenuation Coefficients, and Energy Absorption Coefficients From 10 keV to 100 GeV, NBS Report NSRDS-NBS-29,1990.

[23] R. Nath, L. Anderson, D. Jones,et at. Specification of brachytherapy source strength: A report by Task Group 32 of the American Association of Physicists in Medicine, AAPM Rep. No. 21 ~American Inst. Physics, New York, 1987.

[24] ICRU 38 Dose and Volume Specification for Reporting Intracavitary Therapy in Gynecology, International Commission on Radiation Units and Measurements, 1985.

[25] ICRU 60 Fundamental Quantities and Units for Ionizing Radiation, International Commission on Radiation Units and Measurements, 1998.